# Kinetic Model for Stochastic Heating in the INCA Discharge

*Uwe Czarnetzki*

Institute for Plasma and Atomic Physics, Faculty of Physics and Astronomy,
Ruhr-University Bochum, 44801 Bochum, Germany

**Abstract**

A novel electron heating mechanism based on periodically structured vortex fields induced in a plane was first proposed in 2014 [U. Czarnetzki and Kh. Tarnev, Physics of Plasmas 21, 123508 (2014)]. This theoretical concept has now been realized in an experiment which confirms efficient collisionless heating in such array structures [Ph. Ahr, T.V. Tsankov, J. Kuhfeld, U. Czarnetzki, submitted to Plasma Sources Science and Technology, arXiv:1806.02043v1 (2018)]. The new concept is called "Inductively Coupled Array": INCA. Here, the physical mechanism behind the collisionless (stochastic) heating is investigated by two analytical models. Firstly, the electron heating rate in an array field structure with an exponential spatial decay of the field in the direction perpendicular to the plane is investigated by stochastically averaging single electron trajectories. The approach is similar to the Lieberman model for the classical stochastic heating in standard inductively coupled plasmas. This analysis shows that classical stochastic heating by thermal motion along the vertical direction makes a negligible contribution. However, there is a strong collisonless non-local heating effect in the plane. In conclusion, heating is non-local in the plane but local in the vertical direction. This insight allows a straightforward solution of the collisionless Boltzmann equation which not only confirms the results of the Lieberman model but provides also explicit expressions for the complex conductivity. Based on the conductivity an effective stochastic collision frequency, the complex damping coefficient and the related field penetration of the field into the plasma is calculated. Finally, elastic collisions with neutral background atoms are included in the model and a condition for dominance of stochastic heating over Ohmic heating is derived.

**1. Introduction**

Recently a new way of collisionless electron heating and plasma generation at low pressures was demonstrated experimentally [1]. The so called <u>In</u>ductively <u>C</u>oupled <u>A</u>rray (INCA) discharge consists of a chess-board like regular array of small planar inductive coils driven at radio frequencies (RF) with a fixed phase between the currents in the individual circuits. These coils induce a regularly-structured temporally and spatially varying electric field in the plasma. This vortex field array allows efficient collisionless heating of electrons at pressures in the Pa range and below. The source in the recent experiment consists of 36 coils and has a size of 42 cm x 42 cm. However, extension to $m^2$ size seems to be straight forward, which offers a number of opportunities for application in plasma processing. The basic concept was first proposed in 2014 by Czarnetzki and Tarnev in an initial theoretical investigation [2]. In that work two alternative field structures were discussed: The ortho and the para array. In the ortho array, the currents of all coils are in phase and in the para array currents of adjacent coils have a phase shift of 180°. Resonance structures in velocity space derived from an inspection of the collisionless Boltzmann equation were identified for these periodic field structures. Fields and resonances for the ortho array are shown in Fig. 1. This kind of array is also realized in the recent experiment. The resonances as well as substantial collisionless heating caused by these resonances were confirmed in [2] by a single particle Monte Carlo simulation based on the ergodic principle. The main simplifying assumptions in this simulation were the neglect of any field or plasma inhomogeneity in the dimension perpendicular to the plane of the coils and infinite extension of the plane.

Here an analytical model for the heating is developed. Still infinite extension in the plane and a homogeneous density are assumed but the spatial field decay is included. Two alternative approaches are investigated. Firstly, the Lieberman model for stochastic heating in Inductively Coupled Plasmas (ICPs) [3-5] is extended for the array field structure. The Lieberman model is a simplified version of the far more involving kinetic model developed by Weibel [4]. The essential simplification in the Lieberman model is neglecting the effect of plasma currents on the electric field, i.e. the electric field has a given exponential spatial decay. The Weibel as well as the Lieberman model, which both address collisionless (stochastic) heating in common ICPs, are non-local in the coordinate perpendicular to the coil plane but local in the plane. In contrast, the essential feature of the array discharge is the non-locality in the plane. The analysis based on the Lieberman model shows that the non-locality in the perpendicular coordinate provides a negligible contribution to the stochastic heating in the INCA discharge. The far dominant non-local stochastic heating effects result from the two-dimensional structure in the plane of the array. The general result contains non-trivial integrals but in the local approximation these integrals reduce to explicit analytical expressions.

Vertical locality allows a systematic perturbative solution of the corresponding collisionless Boltzmann equation. The calculation follows the well-known concept of the perpendicular conductivity [7]. The result for the stochastic heating power is identical to the Lieberman model in the limit of locality in the vertical coordinate. Further, the perturbative approach provides explicit expressions for the complex conductivity. Based on the conductivity, the complex damping coefficient is finally derived and the evolution of the electric field amplitude and phase penetrating into the plasma is calculated.

The remaining part of the paper is organized as follows: In section 2 the Lieberman model is applied firstly to the so called ortho array field structure and the heating rate is derived (section 2.1). Subsequently, the same procedure is applied to the so called para array field structure (section 2.2). Both field structures are provided in Appendix A. Further, the classical stochastic heating rate for the mean square field of the ortho array is derived in section 2.3 and finally the classical Ohmic heating rate in section 2.4. A first order perturbative solution of the collisionless Boltzmann equation is found in section 3.1 under the assumption of locality in the vertical direction. This provides not only the heating rate but also the complex conductivity. Based on this result, the complex damping constant is calculated in section 3.2, which allows a description of the propagation of the electromagnetic wave into the plasma. The Boltzmann equation can be extended in a straightforward manner by a relaxation collision operator in order to account in a consistent way for both, Ohmic and stochastic heating (section 3.3). Finally, in section 4 the results are summarized and discussed, including also some outlook for further work on this problem. Numerical values are calculated for a set of standard conditions taken from the experiment. The relevant parameters are summarized in Appendix B. Certain more mathematical parts of the calculation in section 2.1 are moved to Appendix C in order to enhance readability of the main part.

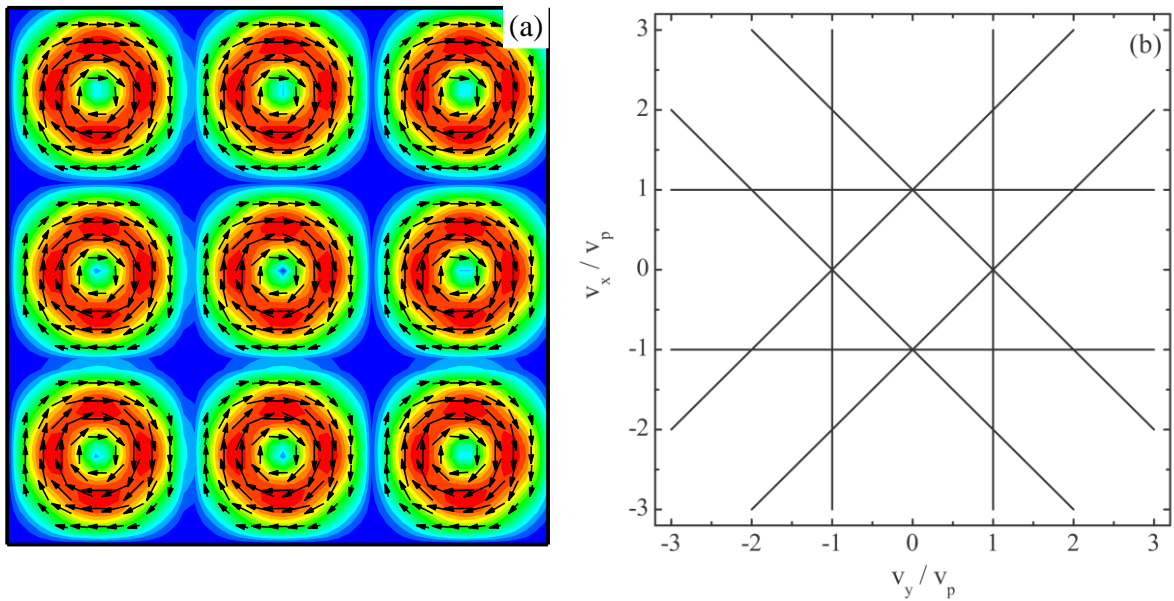

Fig. 1: (a): Electric field structure in the *x,y* plane for the ortho array (the absolute value of the field is encoded by colors with red representing the maximum value $E_0$ and blue zero, instantaneous directions are indicated by the vectors). The size of the figure is $3\Lambda \times 3\Lambda$, where $\Lambda$ is the size of the quadratic cell of an individual coil, and the fields created by the 9 individual coils are clearly visible. (b): Corresponding resonances in velocity space. It should be noted that all resonances are off center. $v_p = \Lambda / T_{RF}$ is the phase velocity of the array for a harmonic RF electric field with period $T_{RF}$. An explicit expression for the field is provided in Appendix A. Figures are from Ref. [2].

## 2. Lieberman Model for the INCA Discharge

In the frame of the Lieberman model collisionless electrons are flowing towards the *x,y*-plane of the ICP coil(s) along the vertical coordinate $(z \geq 0)$ from infinity $(z \to \infty)$, then are reflected by (an infinitely thin) floating potential sheath in the plane of the coils (at $z = 0$), and finally return to infinity (Fig. 2). The electric field is assumed to decay exponentially with distance from the plane of the coils. The aim of the calculation is the determination of the mean power per unit area delivered to the electrons during this transition through the field region. In the original model for standard ICPs, the field is homogenous in the *x,y*-plane and the average is taken over the initial velocity and RF phase. Here, in addition the average is carried out over the *x,y*- (or planar) coordinates. For the electron velocity distribution function $f_e$ an isotropic and homogeneous Maxwellian distribution $f_0$ with a temperature $T_e$ and an electron density $n_0$ is assumed ($k_B$ is the Boltzmann constant, $m_e$ the electron mass, and $v^2 = v_x^2 + v_y^2 + v_z^2$ the squared electron velocity):

$$f_0(v) = n_0 \left( \frac{m_e}{2\pi k_B T_e} \right)^{3/2} e^{-\frac{m_e v^2}{2 k_B T_e}} \qquad (1)$$

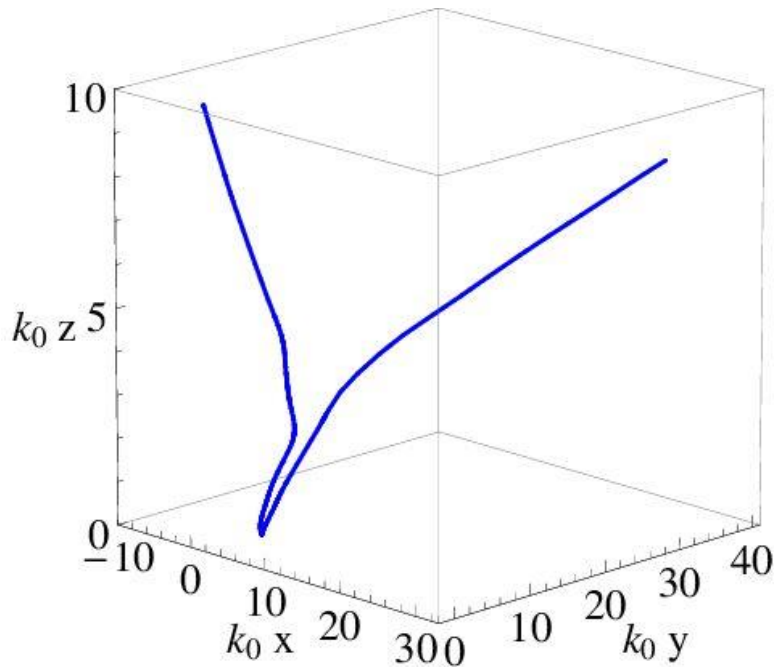

Fig. 2: Coordinate system used in this work with an example of a typical electron trajectory. The trajectory starts at the upper left corner and the particle is reflected at the plane of the coils at the bottom. The vortex electric field induced by these coils is pointing in the *x,y*-direction and is decaying exponentially along the *z*-coordinate with a characteristic normalized length of $s k_0 = 4.2$ with $k_0 = 2\pi / \Lambda$ and $\Lambda = \text{coil cell size}$. The numerical value is taken from the standard conditions listed in Appendix B. Note that the electron experiences a significant gain in the planar velocity while the vertical velocity remains constant.

**2.1 Stochastic Heating in the Ortho Array**

The calculation is carried out first for the so called ortho array where the currents of all coils are in phase [2]. The solution can then be adapted easily to the so called para array structure, where the currents of adjacent coils have a phase shift of 180° [2]. The electric field for the ortho array in the *x,y*-plane is given in Appendix A and also presented in Fig. 1 which shows the amplitude as well as the vortex field vectors pointing in the *x,y*-plane. In addition, in the frame of the Lieberman model, it is assumed that the field decays exponentially in *z*-direction with a characteristic decay length *s*, i.e. perpendicular to the plane of the coils and the induced electric field vectors:

$$\vec{E} = \vec{E}_{ortho}\left(k_0 x, k_0 y, \omega_0 t\right) e^{-\frac{z}{s}}. \tag{2}$$

The periodic structure in space and time is characterized by the wavenumber of the array $k_0 = 2\pi / \Lambda$ and the angular frequency of the field oscillation $\omega_0$. The array structure reproduces itself after a length $\Lambda$. In case of the ortho array this length is identical to the cell size of an individual coil.

The equation of motion for a collisionless electron is ( *e* : elementary charge) :

$$\frac{\partial \vec{v}}{\partial t} = -\frac{e\vec{E}(\vec{r},t)}{m_e} \tag{3}$$

Following the above concept, the change in velocity for an electron coming from and leaving to infinity is calculated:

$$\Delta \vec{v} = -\int_{-\infty}^{+\infty} \frac{e\vec{E}}{m_e} dt \tag{4}$$

The resulting mean power per area delivered to the electrons is:

$$\left\langle \frac{\partial P}{\partial A} \right\rangle = \left\langle -n_e \frac{m_e}{2}\left(\left(\vec{v}_0 + \Delta \vec{v}\right)^2 - \vec{v}_0^2\right) v_{z0} \right\rangle \tag{5}$$

$\vec{v}_0$ is the initial velocity at infinity and $v_{z0} < 0$ the corresponding *z*-component. The average has to be carried out over all initial positions, initial phases, and the initial velocity distribution for all three directions, in total 6 averages.

The *z*-component is not coupled to the field so that $v_z = v_{z0}$ and $z = -v_{z0}|t|$, where $v_{z0} < 0$ is the initial electron velocity in *z*-direction, i.e. perpendicular to the coil plane. Reflection at the surface is considered by the absolute sign ($z \geq 0$). Clearly, equation (3) cannot be integrated exactly due to the in-plane coordinates appearing in the field. The approximation made here is to replace the temporal integral over the actual velocities by $x \approx v_{x0} t$, $y \approx v_{y0} t$, similar to the *z*-component. This approximation requires that the dephasing due to the change in velocity is negligible: $k_0 s |\Delta v_{x,y}/v_{z0}| \ll 2\pi$ or $s/\Lambda |\Delta v_{x,y}/v_{z0}| \ll 1$, where $\Delta v_{x,y}$ is the change in the velocity due to the electric field and $s/v_{z0}$ is the characteristic interaction time. The condition is checked later for self-consistency of the approximation. The initial velocities follow the Maxwell distribution of equation (1). The electric field in the plane of the coils is composed of various vector components in the *x,y*-plane which all have the same general form (Appendix A):

$$\vec{E} = \frac{E_0}{8i} \sum_{\alpha,\beta=0,\pm 1} \vec{A}_{\alpha,\beta}\, e^{i(\omega_{\alpha,\beta} t + \gamma_{\alpha,\beta}) + \frac{v_{z0}}{s}|t|} + c.c.. \tag{6}$$

$E_0$ is the electric field amplitude. The effective frequencies $\omega_{\alpha,\beta} = \vec{k}_{\alpha,\beta}\cdot\vec{v}_0 + \omega_0 = k_0(\alpha v_{x0} + \beta v_{y0}) + \omega_0$ contain already the various resonances in velocity space shown in Fig. 1b by the conditions $\omega_{\alpha,\beta} = 0$ or $v_p = \omega_0/k_0 = \pm(\alpha v_{x0} + \beta v_{y0})$, where $v_p = \omega_0/k_0$ is the phase velocity of the array. Phases $\gamma_{\alpha,\beta} = k_0(\alpha x_0 + \beta y_0) + \omega_0 t_0$ are introduced in order to account for the fact that electrons can start at infinity at different locations $x_0, y_0$ in the plane and at different times $t_0$. After integration and combination of the conjugate complex terms, the velocity change caused by the individual field components reads:

$$\Delta\vec{v}_{\alpha,\beta} = \vec{A}_{\alpha,\beta}\,\frac{eE_0 s v_{z0} \sin(\gamma_{\alpha,\beta})}{2m_e\left(v_{z0}^2 + s^2\omega_{\alpha,\beta}^2\right)}. \tag{7}$$

Note that the resonant velocity directions (parallel to $\vec{k}_{\alpha,\beta}$) and the directions of change in velocity (parallel to $\vec{A}_{\alpha,\beta}$) are always perpendicular since for a vortex field $\vec{A}_{\alpha,\beta} \perp \vec{k}_{\alpha,\beta}$. The condition for negligible dephasing is $s/\Lambda |\Delta v_{x,y}/v_{z0}| \ll 1$. Setting the sine as well as the amplitude factor to one, the frequencies $\omega_{\alpha,\beta}$ to zero, and $m_e v_{z0}^2 = k_B T_e$ maximizes the right hand side of equation (7). A rough condition for the maximum electric field amplitude is then:

$$eE_0 < \frac{2k_B T_e \Lambda}{s^2} \tag{8}$$

For the standard experimental parameters (Appendix B) the critical value is $E_0 < 3.5\,\text{V/cm}$. For smaller skin depths, the critical field would increase strongly. The approximation made here in order to solve the equation of motion can be viewed as a first order approximation.

The phase averages in equation (5) leave only quadratic terms, i.e. all linear and all mixed terms between the different vector components vanish. For the remaining quadratic components a common factor $\langle \sin^2(\gamma_{\alpha,\beta})\rangle = 1/2$ results:

$$\left\langle (\vec{v}_0 + \Delta\vec{v})^2 - \vec{v}_0^2 \right\rangle_{\gamma_{\alpha,\beta}} = \sum_{\alpha,\beta=0,\pm 1}\left\langle \Delta v_{\alpha,\beta}^2 \right\rangle_{\gamma_{\alpha,\beta}} = \frac{e^2 E_0^2 s^2}{8m_e^2}\sum_{\alpha,\beta=0,\pm 1}\left|\vec{A}_{\alpha,\beta}\right|^2 \frac{v_{z0}^2}{\left(v_{z0}^2 + s^2 \omega_{\alpha,\beta}^2\right)^2}. \tag{9}$$

This expression can now be averaged over all negative initial $z$-velocities in equation (5) assuming Maxwell distributed electrons (equation (1)):

$$\left\langle \frac{\partial P}{\partial A} \right\rangle_{ortho}^{(L)} = \left\langle \frac{n_0 e^2 E_0^2 s^2}{16 m_e} \sqrt{\frac{m_e}{2\pi k T_e}} \sum_{\alpha,\beta=0,\pm 1} \frac{1}{\alpha^2+\beta^2} \int_0^\infty \frac{v_{z0}^2 e^{-\frac{m_e v_{z0}^2}{2k_B T_e}}}{\left(v_{z0}^2 + s^2 \omega_{ij}^2\right)^2} v_{z0}\, dv_{z0} \right\rangle_{v_{x0},v_{y0}}$$

$$= \left\langle \frac{n_0 e^2 E_0^2 s^2}{32 m_e} \sqrt{\frac{m_e}{2\pi k T_e}} \sum_{\alpha,\beta=0,\pm 1} \frac{1}{\alpha^2+\beta^2} \int_0^\infty \frac{\varepsilon\, e^{-\varepsilon}}{\left(\varepsilon + \varepsilon_{\alpha\beta}\right)^2} d\varepsilon \right\rangle_{v_{x0},v_{y0}} \quad (10)$$

For convenience the transformation $-v_{z0} \to v_{z0}$ has been applied in the first line of the above equation. Further, $\left|\vec{A}_{\alpha,\beta}\right|^2 = 1/\left(\alpha^2+\beta^2\right)$ has been used. $\varepsilon_{\alpha,\beta} = m_e s^2 \omega_{\alpha,\beta}^2 /(2k_B T_e)$ are the resonant energies normalized to the thermal energy of the electrons. The superscript ($L$) indicates that this result is obtained in the frame of the Lieberman model. Subsequently the same quantity will be calculated also by a perturbative solution of the collisionless Boltzmann equation which will be indicated by a superscript ($B$). The integral is identical to the integral appearing in the original Lieberman model [3]:

$$H\left(\varepsilon_{\alpha,\beta}\right) = \int_0^\infty \frac{\varepsilon\, e^{-\varepsilon}}{\left(\varepsilon + \varepsilon_{\alpha,\beta}\right)^2} d\varepsilon = \left(1+\varepsilon_{\alpha,\beta}\right) e^{\varepsilon_{\alpha,\beta}} E_1\left(\varepsilon_{\alpha,\beta}\right) - 1. \quad (11)$$

The standard integral $E_1(x)$ is related to the more common exponential integral $Ei(x)$ [8]:

$$E_1(x) = \int_x^\infty \frac{e^{-t}}{t} dt = -Ei(-x). \quad (12)$$

The sum in equation (10) is carried out by considering that there are effectively only 2 different terms representing parallel and diagonal resonances in velocity space and corresponding parallel and diagonal fields (with respect to the coordinate axes). Both have a multiplicity of 4 but the former have amplitudes 1 and the latter amplitudes ½ (Appendix A):

$$\left\langle \frac{\partial P}{\partial A} \right\rangle_{ortho}^{(L)} = \frac{n_0 e^2 E_0^2 s^2}{8 m_e} \sqrt{\frac{m_e}{2\pi k T_e}} \left( \left\langle H(\varepsilon_\parallel) \right\rangle_{v_{x0}} + \frac{1}{2} \left\langle H(\varepsilon_{//}) \right\rangle_{v_{x0},v_{y0}} \right). \quad (13)$$

The normalized axial and diagonal energies are $\varepsilon_\parallel = m_e s^2 (k_0 v_{x0} - \omega_0)^2 /(2k_B T_e)$ and $\varepsilon_{//} = m_e s^2 (k_0 (v_{x0}+v_{y0}) - \omega_0)^2 /(2k_B T_e)$, respectively. Two dimensionless integrals can be defined which both depend on two dimensionless parameters $\psi = s k_0 = 2\pi s/\Lambda$ and $\chi = \sqrt{\varepsilon_p/(k_B T_e)}$, where $\varepsilon_p = m_e v_p^2/2$ is the characteristic phase energy of the array:

$$G_1(\chi,\psi) = \left\langle H(\varepsilon_\parallel) \right\rangle_{v_{x0}} = \frac{\psi}{\sqrt{\pi}} \int_{-\infty}^\infty H\left(\psi^2 (u-\chi)^2\right) e^{-u^2} du \approx \frac{\pi}{2} e^{-\chi^2}, \quad (14)$$

$$G_2(\chi,\psi) = \langle H(\varepsilon_{//}) \rangle_{v_{x0},v_{y0}} = \frac{\psi}{\pi} \int_{-\infty}^{\infty} \int_{-\infty}^{\infty} H(\psi^2(u+w-\chi)^2) e^{-u^2-w^2} du\, dw$$

$$= \frac{\psi}{\sqrt{2\pi}} \int_{-\infty}^{\infty} H(\psi^2(u-\chi)^2) e^{-\frac{1}{2}u^2} du \quad (15)$$

$$\approx \frac{\pi}{2^{3/2}} e^{-\chi^2/2}$$

In equation (15), the transformation $u+w=p$ and $u-w=q$ with $du\,dw = dp\,dq/2$ has been applied. This allows to single out the integral over $q$ which yields $\sqrt{2\pi}$. In the remaining integral change of the denotation $p \to u$ is used for convenience when comparing it with $G_1$. Both integrals do not have an exact analytical solution and are solved numerically in Fig. 3. The approximate formulas valid for $\psi \gg 1/\sqrt{2}$ are derived in Appendix C. The approximate expressions exhibit that $G_{1,2} \to 0$ for $k_B T_e \ll \varepsilon_p$ and $G_{1,2} \to$ constant for $k_B T_e \gg \varepsilon_p$. Clearly stochastic heating in the array is a hot plasma effect. Further, the integrals are effectively independent of the parameter $\psi$, the ratio of the skin depth to the cell size, if $\psi \gg 1/\sqrt{2}$, which ensures dominance of the in-plane stochastic heating over the classical vertical stochastic heating. A more detailed discussion of the contribution of classical stochastic heating follows in section 2.3. Vertical locality is also of key importance in connection with the solution of the collisionless Boltzmann equation in section 3. The physical reason for the slightly different behavior of the axial and diagonal resonances is by the simple fact that the distance between the individual vortex fields along the diagonal is $\sqrt{2}$ larger than along the coordinate axes. The explicit analytical expressions derived for the limiting case $\psi \gg 1/\sqrt{2}$ are also reproduced in section 3.1 by a first order perturbative solution of the collisionless Boltzmann equation under the assumption of vertical locality.

The final result is now:

$$\left\langle \frac{\partial P}{\partial A} \right\rangle_{ortho}^{(L)} = \frac{n_0 e^2 E_0^2 s}{8\sqrt{\pi}\, m_e \omega_0} G_{ortho}^{(L)}(\chi,\psi), \quad (16)$$

with

$$G_{ortho}^{(L)}(\chi,\psi) = \chi\left(G_1(\chi,\psi) + \frac{1}{2} G_2(\chi,\psi)\right)$$

$$\approx \frac{\pi}{2} \chi\left(e^{-\chi^2} + \frac{1}{2^{3/2}} e^{-\frac{\chi^2}{2}}\right) \quad . \quad (17)$$

It should be noted that the dependence of the mean power per area by $n_0 e^2 E_0^2 s/(m_e \omega_0)$ times a function depending on the dimensionless control parameters $\psi$ and $\chi$ could have been derived right from the start by a dimensional analysis. Not surprising, the same general relation appears also in all other cases in this work.

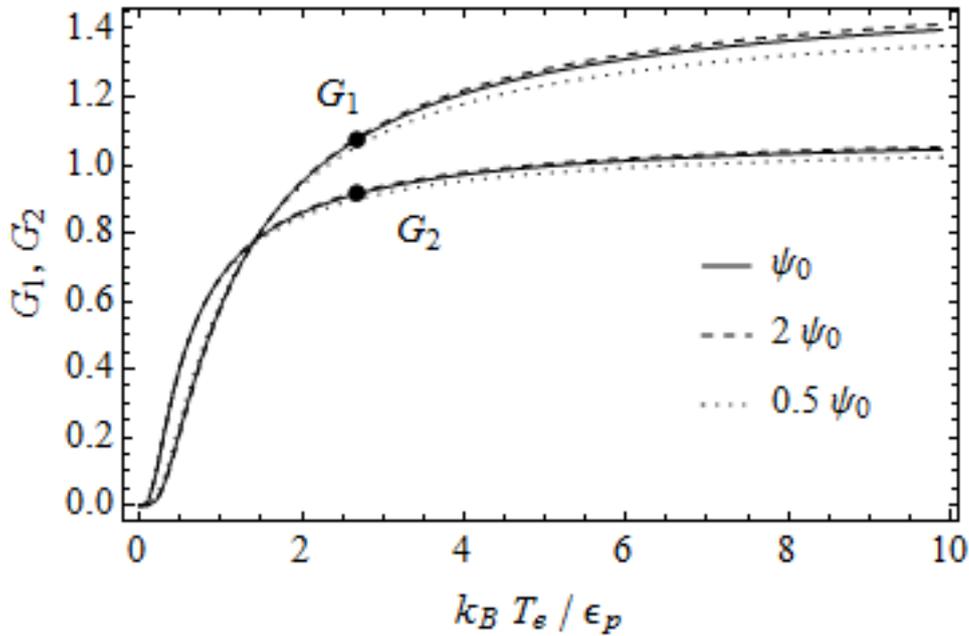

Fig. 3: The functions $G_1$ and $G_2$ as a function of the electron temperature normalized to the phase velocity energy of the array for three different values of the parameter $\psi = s k_0 = 2\pi s / \Lambda$ with $\psi_0 = 4.2$. Dots represent standard conditions (Appendix B).

The function $G_{ortho}^{(L)}$ is shown in Fig. 4. The superscript (L) indicates that the function was calculated based on the Lieberman model. Like the individual functions $G_1$ and $G_2$, representing the parallel and the diagonal resonances, respectively, also the effective function $G$ vanishes for small electron temperatures. For very large electron temperatures the interaction time in the heating zone becomes shorter and the efficiency decays by the inverse thermal speed. Since already the functions $G_1$ and $G_2$ are insensitive to variations in $\psi$ the same characteristic is found for $G_{ortho}^{(L)}$. The function has a maximum of $G_{max} \approx 0.98$ at about $k_B T_e \approx 1.6 \varepsilon_p$ which corresponds to about 2.1 eV for the standard conditions. The heating efficiency remains within 80 % of the maximum value for $0.7 < k_B T_e / \varepsilon_p < 5$. This covers a rather broad range of temperatures so that with a proper choice of $\varepsilon_p$ stochastic heating remains efficient for all reasonable variations in pressure and gas type, under the standard conditions from 0.9 eV to 6.5 eV. The standard value of $k_B T_e = 3.5 \text{eV}$ is quite in the center of this range.

At the experimental standard conditions (Appendix B) a power of about $P = 430 W$ is delivered to the plasma. There $G_{ortho}^{(L)} \approx 0.93$ and so an electric field of $E_0 = 4.9 \text{V/cm}$ would be required according to equation (16) and considering the total array size of $A = 0.090 \text{m}^2$. The electric field in vacuum at the surface of the $d = 4 \text{mm}$ quartz window in front of a single ideal two-turn coil for the experimental peak current of $I = 12 \text{A}$ is calculated as $E_0 = 5.4 \text{V/cm}$ by combining Biot-Savart's law with the induction law. Therefore, within the experimental uncertainties of about 20 % the field is very reasonable. Note that the skin depth enters only linearly, i.e. as a factor representing the effective heating depth or volume. The cell size $\Lambda$ has only a very weak influence via the function $G_{ortho}^{(L)}$ which effectively is almost a constant in the relevant parameter range of electron temperatures for a proper choice of $\Lambda$.

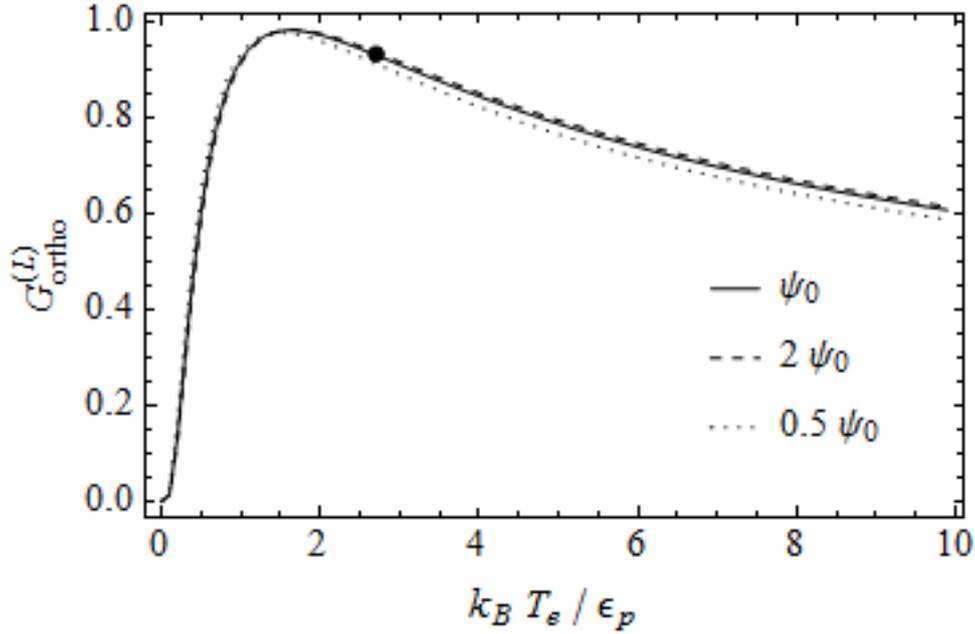

Fig. 4: The effective function $G_{ortho}^{(L)}$, describing the electron temperature dependence of the stochastic heating in the ortho array. $G_{ortho}^{(L)}$ is shown for three different values of the parameter $\psi = s k_0 = 2\pi s / \Lambda$ with $\psi_0 = 4.2$. The dot represents standard conditions (Appendix B).

Finally, the mean energy gained per electron can be calculated as the above power per area divided by the flux of the inflowing electrons:

$$\langle \Delta \varepsilon \rangle_{otho} = \frac{e^2 E_0^2}{4 m_e \omega_0^2} \psi \, \chi \, G_{ortho}^{(T)}(\chi, \psi). \tag{18}$$

The first term is the mean energy of a free oscillating electron. The subsequent dimensionless factor has a similar behavior as the $G$ function but due to the additional factor $\chi$ it peaks already at about $1/\chi^2 = k_B T_e / \varepsilon_p \approx 0.75$ with a peak value of $0.95 \psi$. Naturally, it decays like $\chi^2 = \varepsilon_p / (k_B T_e)$ for larger values of the temperature. For the standard conditions the entire factor is about 2.4 and for a field amplitude of $E_0 = 4.9\,\text{V/cm}$ one calculates $\langle \Delta \varepsilon \rangle_{ortho} \approx 3.5\,\text{eV}$, i.e. accidentally the same value as $k_B T_e$ which represents a significant gain.

### 2.2 Stochastic Heating in the Para Array

The para array has the same chess-board arrangement of coils but currents and corresponding induced fields of adjacent coils are 180° out of phase. The electric field of the para array is also listed in Appendix A. One of the main consequences is that now the fields of adjacent coils interfere constructively and not destructively, as in the ortho array, and the overall field structure is slightly altered with fewer voids [2]. As a quantitative measure for comparison one can take the spatial mean square field which is $3/8 E_0^2$ in case of the ortho array and $1/2 E_0^2$ in case of the para array, i.e. a ratio of 4/3 in favor of the para array.

Inspecting the explicit form of the electric field structure (Appendix A) reveals immediately that the general problem is identical to the ortho array but without the parallel resonances, i.e. without $G_1$ contributing in the $G$ function. Further, the para array does not have the factor ½ for the diagonal field amplitudes which leads in the final result for the power per area to a factor $2^2 = 4$ when applying the formula derived for the ortho array. This factor is included in the $G$ function which then changes to $G_{para}^{(L)}(\chi,\psi) = 2\chi G_2(\chi,\psi)$. The function is shown in Fig. 5. Naturally, the behavior is very similar to the ortho array, especially in respect to the already discussed dependence on $\psi$ and the related locality in the vertical coordinate. Therefore, the overall performance in the experiment can be expected to be similar too. However, the $G$ function of the para array is generally slightly larger which can be attributed to the more effective field distribution and the related factor 4/3 discussed above.

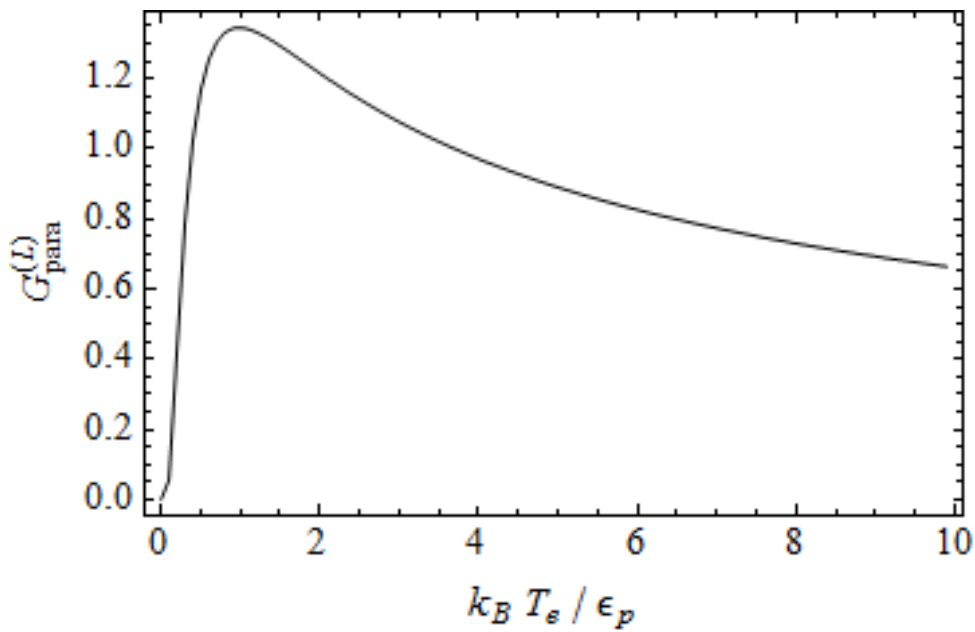

Fig. 5: The function $G_{para}^{(L)}$, describing the electron temperature dependence of the stochastic heating in the para array for $\psi_0 = 4.2$, representing the standard conditions.

### 2.3 Classical Stochastic Heating in the Ortho Array

The new non-local heating discussed above applies for large locality parameters $\psi \gg 1$. The opposite limit $\psi \ll 1$ leads to classical stochastic heating. As is shown in Appendix C, in this case the electron temperature dependent $G$ function becomes:

$$G_{ortho}^{(L)} \to G_{csh}^{(L)}(\psi\chi) = \frac{3}{2}\psi\chi H\left((\psi\chi)^2\right). \tag{19}$$

Naturally, the preceding amplitude factor is the same as in equation (16). The corresponding total formula for the power deposition per unit area is then identical to the result obtained by Lieberman [3] for the classical stochastic heating in ICPs except for a factor 3/8. This factor follows indeed from the spatial average of the squared electric field $\left\langle E_{ortho}^2 \right\rangle_{x,y} = 3/8 E_0^2 \exp(-2z/s)\cos^2(\omega_0 t)$ and simply reflects the fact that the classical case is local in the in-plane coordinates.

The transition to the classical case can be made by letting $\Lambda \to \infty$, i.e. very large individual coils, which results in $\psi \to 0$ and $\chi \to \infty$ with the product $\psi\chi = \sqrt{s^2 \omega_0^2 m_e / (2k_B T_e)}$ remaining finite since there $\Lambda$ cancels out. Actually the product replaces $\Lambda$ as the scaling length by $s$. Fig. 6 shows the characteristics of $G_{csh}^{(L)}$. However, one should not apply this result under the standard conditions of the ortho array since apparently for these conditions $\psi \gg 1$. Nevertheless, in Fig. 6 also the value of $G_{csh}^{(L)}$ for standard conditions is shown. In this case, one should focus the attention not so much on the value of $G_{csh}^{(L)}$ but on the value of the effective variable. Apparently, this variable $k_B T_e / (\varepsilon_p \psi^2)$ has a very low and therefore quite improper value for stochastic heating. The downscaling is clearly caused by the high value of the locality parameter $\psi^2 = 16$, which reduces the effective variable by more than an order of magnitude. Obviously, with the other parameters remaining constant, classical stochastic heating would be highly effective for $\psi \leq 1$. In conclusion, the insight provided by the above discussion is in highlighting the role of the locality parameter $\psi$ and of the wide applicability of the general result that clearly contains both important limits.

The role of classical stochastic heating can also be estimated by the electron residence time within the skin depth. In accordance with Ref. [3] one finds for the ratio of the residence time to the RF period $\tau_s / T_{RF} = 2\psi\chi / \sqrt{\pi} \approx 2.7$ (numerical value for the standard conditions). Therefore, on average the electron spends much more than one RF period in the field region. For the classical stochastic heating to be efficient, the residence time should be shorter than the RF period. This would be the case particularly for substantially lower RF frequencies in the MHz range which are known to be more efficient in providing stochastic heating [9].

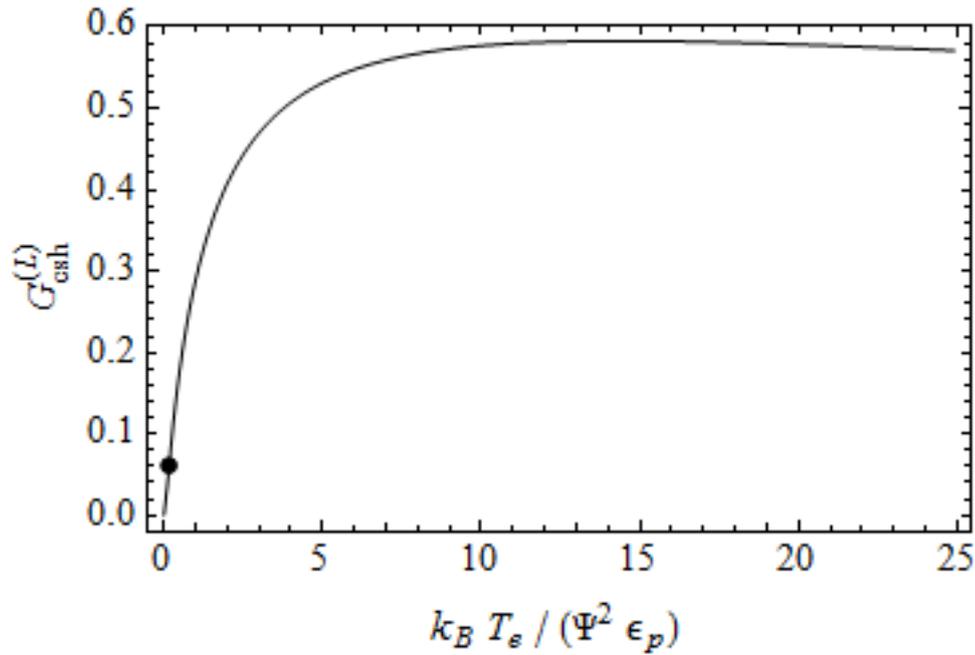

Fig. 6: $G_{cst}^{(L)}$ function showing the electron temperature dependence of the classical stochastic heating in the ortho array. The dot represents standard conditions (Appendix B).

**2.4 Ohmic heating in the Ortho Array**

The basic principles of Ohmic heating are locality and collisionality with background atoms. In order to calculate the Ohmic heating power one has to add to the right hand side of the equation of motion (3) a friction term $-\nu_m \vec{v}$ with the elastic collision frequency for momentum transfer $\nu_m$. In this case $\vec{v}$ and $\nu_m$ denote no longer the velocity and collision rate of an individual electron but describe the ensemble average values. The resulting velocity and the related current density $\vec{j} = -e n_0 \vec{v} = \sigma_{Ohm} \vec{E}$ are proportional to the local electric field. The complex Ohmic conductivity is:

$$\sigma_{Ohm} = \sigma_0 \frac{1}{\zeta_m + i} = \sigma_0 \frac{\zeta_m - i}{1 + \zeta_m^2}, \qquad (20)$$

where the dimensionless collision parameter is $\zeta_m = \nu_m / \omega_0$. This parameter as well as $\sigma_0$ will reappear in the stochastic conductivity derived in section 3 ($\omega_{pe}$: electron plasma frequency, $c$: speed of light, $\mu_0$: vacuum permeability):

$$\sigma_0 = \frac{n_0 e^2}{m_e \omega_0} = \frac{\omega_{pe}^2}{\mu_0 c^2 \omega_0} \qquad (21)$$

The mean power per volume $\langle \partial P / \partial V \rangle = \langle \vec{j} \cdot \vec{E} \rangle_{T_{RF}}$ is proportional to the square of the local field. Only the real part of the conductivity contributes and the temporal average over the harmonic function yields a factor ½. The power per area is calculated by integrating the squared electric field over $z$ from zero to infinity which gives a factor $s/2$. In addition, the factor 3/8 has to be included for the average over the plane as discussed above. The final result can be cast into a similar form as obtained for the stochastic heating:

$$\left\langle \frac{\partial P}{\partial A} \right\rangle_{ohm} = \frac{n_e e^2 E_0^2 s}{8\sqrt{\pi}\, m_e \omega_0} G_{ohm}\left(\frac{\nu_m}{\omega_0}\right), \qquad (22)$$

where

$$G_{ohm}(\zeta_m) = \frac{3\sqrt{\pi}}{4} \frac{\zeta_m}{1+\zeta_m^2} \qquad (23)$$

In this case, there is of course no dependence on the electron temperature and the $G$ function defined here depends only on the ratio between the elastic collision frequency and the RF frequency. It is remarkable that $G_{ohm} < 1$ with the maximum of 0.66 at $\zeta_m = 1$. Therefore, it seems that even at the most favorable collisionality, the Ohmic contribution to the electron heating is smaller than the stochastic contribution in the ortho array. However, this simple reasoning is limited by the fact that in order for the stochastic heating to be effective, the electron mean free path must be larger than the size of a single coil $v_{th}/\nu_m > \Lambda$. For the standard conditions the thermal speed of the electrons is $v_{th} \approx 1.3 \cdot 10^6$ m/s and $\nu_m = 3 \cdot 10^7 s^{-1}$. The ratio of $v_{th}/\nu_m = 4.3$ cm is in fact close to the cell size of $\Lambda = 5 cm$. Indeed, the experimental threshold pressure for stochastic heating in Argon is around the standard condition pressure of $p = 1.0 Pa$. At this limit the Ohmic $G$ factor is $G_{ohm} \approx 0.42$, i.e. the Ohmic heating can be estimated to contribute only 30 % to the total heating. For Neon, however, collisionality is much lower and the pressure can be an order of magnitude higher [10,11]. The inconsistency in the above estimate is apparently in treating Ohmic and stochastic heating separately. A consistent treatment of both effects is provided in section 3.3 under the assumption of vertical locality.

## 3. Boltzmann Model for the INCA Discharge
### 3.1 Perturbative Solution of the Collisionless Boltzmann Equation

A comprehensive treatment of the problem of stochastic heating would solve the collisionless Boltzmann equation for the electron velocity distribution function together with the wave equation describing the propagation of the electromagnetic field. This was done by Weibel for the classical stochastic heating in ICPs. Although this model is only local in the plane, it leads already to rather extensive calculations with complex results. Here, the insight of vertical locality provided by the Lieberman model of the array greatly simplifies the calculation by effectively decoupling the collisionless Boltzmann equation from the wave equation. As usual, the equation is solved perturbatively to first order: $f(\vec{r},\vec{v},t) = f_0(v) + f_1(\vec{r},\vec{v},t)$, where $f_0$ is the isotropic Maxwell distribution of equation (1). The smallness parameter justifying the perturbative approach is the electric field. Naturally, in a vertically local approach the skin depth does not appear but the requirement here is $eE_0 \ll k_0 k_B T_e$. For the standard conditions accidentally almost the same value as in section 2.1 results.

The linearized collisionless Boltzmann equation reads (indices at the nabla operators indicate derivatives with respect to space "r" or velocity "v"):

$$\frac{\partial f_1}{\partial t} + \vec{v} \cdot \nabla_r f_1 \approx \frac{e}{m_e} \vec{E} \cdot \nabla_v f_0(v) . \tag{24}$$

The solution is obtained by Fourier transformation [2]:

$$f_1(\vec{r},\vec{v},t) = \frac{eE_0}{8k_B T_e} \sum_{\alpha,\beta=0,\pm1} \frac{\vec{A}_{\alpha,\beta} \cdot \vec{v} f_0}{\omega_0 + k_0(\alpha v_x + \beta v_y)} e^{ik_0(\alpha x + \beta y) + i\omega_0 t} + c.c. \tag{25}$$

Here the field structure for the ortho array (Appendix A) is already inserted. Since in the vortex field, the $k$ vector is perpendicular to the electric field, one can write alternatively:

$$f_1(\vec{r},\vec{v},t) = \frac{eE_0}{8k_B T_e} \sum_{\alpha,\beta=0,\pm1} \frac{A_{\alpha,\beta} v_{\|\alpha,\beta} f_0}{\omega_0 + k_{\alpha,\beta} v_{\perp\alpha,\beta}} e^{ik_0(\alpha x + \beta y) + i\omega_0 t} + c.c. \tag{26}$$

Here $v_{\|\alpha,\beta}$ is the velocity component parallel to the (dimensionless) electric field vector $\vec{A}_{\alpha,\beta}$ and $v_{\perp\alpha,\beta}$ the velocity component parallel to the associated wave vector $\vec{k}_{\alpha,\beta}$. Note that $k_{\alpha,\beta}/k_0 = \sqrt{\alpha^2+\beta^2} = 1$ or $\sqrt{2}$, respectively, with the former value applying for one of the indices being zero (axial fields) and the latter value for both indices being non-zero (diagonal fields). A straight forward calculation shows that there is no space charge associated with $f_1$ or any higher order. However, the current density is non-zero:

$$\begin{aligned}
\vec{j} &= -e \int \vec{v} f_1 d^3v \\
&= \frac{-e^2 E_0}{8k_B T_e} \sum_{\alpha,\beta=0,\pm1} \int \frac{A_{\alpha,\beta} v_{\|\alpha,\beta} \vec{v} f_0}{\omega_0 + k_{\alpha,\beta} v_{\perp\alpha,\beta}} d^3v \, e^{ik_0(\alpha x + \beta y) + i\omega_0 t} + c.c. \\
&= \frac{e^2}{i k_B T_e} \sum_{\alpha,\beta=0,\pm1} \int \frac{v_{\|\alpha,\beta}^2 f_0}{\omega_0 + k_{\alpha,\beta} v_{\perp\alpha,\beta}} d^3v \, \frac{E_0}{8i} \vec{A}_{\alpha,\beta} e^{ik_0(\alpha x + \beta y) + i\omega_0 t} + c.c. \\
&= \frac{e^2 n_0}{i k_B T_e} \sum_{\alpha,\beta=0,\pm1} \int_{-\infty}^{\infty} \frac{e^{-\frac{m_e v_{\perp\alpha,\beta}^2}{2k_B T_e}}}{\omega_0 + k_{\alpha,\beta} v_{\perp\alpha,\beta}} dv_{\perp\alpha,\beta} \, \frac{E_0}{8i} \vec{A}_{\alpha,\beta} e^{ik_0(\alpha x + \beta y) + i\omega_0 t} + c.c. \\
&= \frac{e^2 n_0}{i m_e \omega_0} \sum_{\alpha,\beta=0,\pm1} \xi_{\alpha,\beta} \frac{1}{\sqrt{\pi}} \int_{-\infty}^{\infty} \frac{e^{-w^2}}{w + \xi_{\alpha,\beta}} dw \, \frac{E_0}{8i} \vec{A}_{\alpha,\beta} e^{ik_0(\alpha x + \beta y) + i\omega_0 t} + c.c. \\
&= \frac{e^2 n_0}{i m_e \omega_0} \sum_{\alpha,\beta=0,\pm1} \xi_{\alpha,\beta} Z(-\xi_{\alpha,\beta}) \frac{E_0}{8i} \vec{A}_{\alpha,\beta} e^{ik_0(\alpha x + \beta y) + i\omega_0 t} + c.c. .
\end{aligned} \tag{27}$$

Here $\xi_{\alpha,\beta} = \chi/\sqrt{\alpha^2+\beta^2} = \chi$ or $\chi/\sqrt{2}$, respectively, and $Z(-\xi_{\alpha,\beta})$ is the plasma dispersion function [12, 13]. $\chi = \sqrt{\varepsilon_p/(k_B T_e)}$ as in section 2. Since the argument is real, the plasma dispersion function takes a particularly simple form:

$$Z(-\xi_{\alpha,\beta}) = -Z^*(\xi_{\alpha,\beta}) = i\sqrt{\pi}\, e^{-\xi_{\alpha,\beta}^2} + 2F(\xi_{\alpha,\beta}), \tag{28}$$

with the Dawson $F$ function (also sometimes denoted as $D_+$ or $S$) defined as in [8]:

$$F(\xi_{\alpha,\beta}) = e^{-\xi_{\alpha,\beta}^2} \int_0^{\xi_{\alpha,\beta}} e^{w^2}\, dw. \tag{29}$$

One can now define a general stochastic conductivity, with $\sigma_0$ as in section 2.4:

$$\sigma_{sto}(\xi_{\alpha,\beta}) = i\sigma_0 \xi_{\alpha,\beta} Z^*(\xi_{\alpha,\beta}) = \sigma_0 \xi_{\alpha,\beta}\left(\sqrt{\pi}\, e^{-\xi_{\alpha,\beta}^2} - i2F(\xi_{\alpha,\beta})\right). \tag{30}$$

Alternatively, one can distinguish axial and diagonal expressions:

$$\sigma_\| = i\sigma_0 \chi Z^*(\chi), \quad \sigma_{//} = i\sigma_0 \frac{\chi}{\sqrt{2}} Z^*\left(\frac{\chi}{\sqrt{2}}\right). \tag{31}$$

This clearly highlights the complex characteristic of the conductivity, causing a phase shift between electric field and current density, and the stochastic nature of the electron heating process in the array. A graphical presentation of the conductivity is provided in section 3.3, where also elastic collisions are included. The mean power per volume is:

$$\begin{aligned}\langle \vec{j}\cdot\vec{E}\rangle &= \sum_{\alpha,\beta=0,\pm 1} 2\,\mathrm{Re}(\sigma_{sto})\left|\frac{E_0 \vec{A}_{\alpha,\beta}}{8}\right|^2 \\ &= \frac{n_0 e^2 E_0^2}{32\, m_e \omega_0} \sum_{\alpha,\beta=0,\pm 1} \frac{\sqrt{\pi}\, \xi_{\alpha,\beta}\, e^{-\xi_{\alpha,\beta}^2}}{\alpha^2+\beta^2} \\ &= \frac{n_0 e^2 E_0^2}{4\sqrt{\pi}\, m_e \omega_0}\frac{\pi}{2}\chi\left(e^{-\chi^2} + \frac{1}{2^{3/2}}e^{-\frac{\chi^2}{2}}\right)\end{aligned} \tag{32}$$

Assuming the same exponential decay of the field along the z-coordinate as in the Lieberman model and integrating along $z$ yields the mean power per area by multiplying the above result with $s/2$:

$$\left\langle\frac{\partial P}{\partial A}\right\rangle^{(B)}_{ortho} = \frac{n_0 e^2 E_0^2 s}{8\sqrt{\pi}\, m_e \omega_0} G^{(B)}_{ortho}(\chi), \tag{33}$$

where

$$G^{(B)}_{ortho}(\chi) = \frac{\pi}{2}\chi\left(e^{-\chi^2} + \frac{1}{2^{3/2}}e^{-\frac{\chi^2}{2}}\right). \tag{34}$$

This is indeed the identical result as in the Lieberman model (equation (17)) in the limit of vertical locality $\left(\psi \gg 1/\sqrt{2}\right)$. Naturally, similar agreement is found for the para array.

**3.2 Effective Stochastic Collision Frequency and Complex Damping Coefficient**

Definition of an effective stochastic collision frequency based on the above complex conductivities is desirable in order to have a single and less abstract parameter highlighting the efficiency of the stochastic heating. One tempting way is a direct comparison with the classical Ohmic result (equation (20)) in order to define an effective stochastic collision frequency $\zeta_{eff} = \nu_{eff} / \omega_0$. Indeed, both expressions are proportional to the dimensional factor $\sigma_0$. The imaginary parts are both negative, reflecting the inertia of the electron due to its finite mass. The cases $\xi_{\alpha,\beta} \to \infty$ and $\nu_{eff}/\omega_0 \to 0$ correspond in the sense that in this case the real part vanishes and the imaginary part becomes 1 on both sides. In the opposite case $\xi_{\alpha,\beta} \to 0$ and $\nu_{eff}/\omega_0 \to \infty$ both real parts vanish, but the imaginary part of the Ohmic side vanishes too while the stochastic part converges to -1. In addition, the maximum of the real part of the Ohmic conductivity is $1/2$ (at $\nu_{eff}/\omega_0 = 1$) but for the stochastic expression it is $\sqrt{\pi/(2e)} \approx 0.76 > 1/2$ (at $\xi_{\alpha,\beta} = 1/\sqrt{2}$). Therefore, the functional regions do not match and a direct comparison is questionable. Nevertheless, a reasonable definition can be made by the phase between the electric field and the current density $\Delta\theta = \arctan\left(\operatorname{Im}(\sigma)/\operatorname{Re}(\sigma)\right)$:

$$\zeta_{eff} = \frac{\nu_{eff}}{\omega_0} = -\frac{\operatorname{Re}(\sigma_{sto})}{\operatorname{Im}(\sigma_{sto})} = \frac{\sqrt{\pi}}{2} \frac{e^{-\xi_{\alpha,\beta}^2}}{F(\xi_{\alpha,\beta})} \approx \frac{\sqrt{\pi}}{2\xi_{\alpha,\beta}} = \sqrt{\frac{\pi k_B T_e (\alpha^2 + \beta^2)}{4\varepsilon_p}} \qquad (35)$$

The approximate expressions on the right are the leading order for $\xi_{\alpha,\beta} \ll 1$. Fig. 7 shows that this simple form represents reasonably well the behavior in the relevant electron temperature range. Therefore, in the above definition $\zeta_{eff}$ increases with the square root of the electron temperature and $\nu_{eff} \sim v_{th}/\Lambda$, which might have been anticipated right from the start. Apparently, for standard conditions the effective collision frequency is of the same order as the angular RF frequency and of the same order as the elastic collision frequency for momentum transfer in Argon at a pressure of about 5 Pa. In the experiment, operation in the stochastic mode has been demonstrated down to pressures of 0.1 Pa [1]. At this low pressure, the electron temperature is almost a factor of two higher and the effective stochastic collision frequency is increasing accordingly while the elastic collision frequency reaches totally insignificant values. Although comparison with Ohmic heating and the related elastic collisions cannot be stretched too far, it should be noted nevertheless that the typical value of the effective stochastic collision frequency of $\zeta_{eff} \sim 1$ is of the same order as the optimum value for Ohmic heating as discussed in section 2.4. This can be taken as a further indication of the effectiveness of the stochastic heating process.

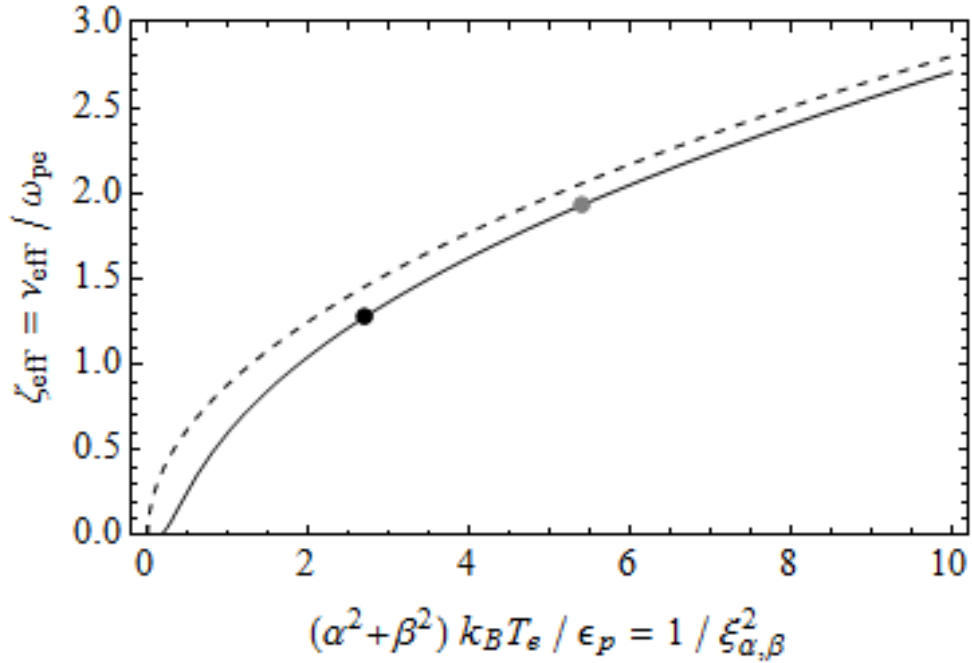

Fig. 7: Effective stochastic collision frequency as defined in equation (35). Solid line: Exact expression, Dashed line: Approximation. Dots represent standard conditions (Appendix B): $\alpha^2 + \beta^2 = 1$ (Black), $\alpha^2 + \beta^2 = 2$ (Gray).

More importantly, the stochastic conductivities are very useful expressions in connection with the complex damping coefficient $\kappa = \kappa' + i\kappa''$ describing the penetration of the RF wave into the plasma, i.e. $E \sim \exp(-\kappa z)$ [3]:

$$\kappa^2 = i\mu_0 \omega_0 \sigma \tag{36}$$

Alternatively, one could use instead of $\kappa$ the complex refractive index $n = c\kappa/\omega_0$ or the complex dielectric constant $\varepsilon = \kappa^2/(\omega_0^2 \mu_0)$ but the damping constant seems to be the more practical choice here since it relates directly to the skin depth $s = 1/\kappa'$.

Denoting the complex conductivity as $\sigma = \sigma' + i\sigma''$ and solving the coupled equations (36) for the real and imaginary parts yields:

$$\begin{aligned} \kappa' &= \sqrt{\frac{\mu_0 \omega_0}{2}} \sqrt{-\sigma'' + \sqrt{\sigma''^2 + \sigma'^2}} \\ \kappa'' &= \sqrt{\frac{\mu_0 \omega_0}{2}} \frac{\sigma'}{\sqrt{-\sigma'' + \sqrt{\sigma''^2 + \sigma'^2}}} \end{aligned} \tag{37}$$

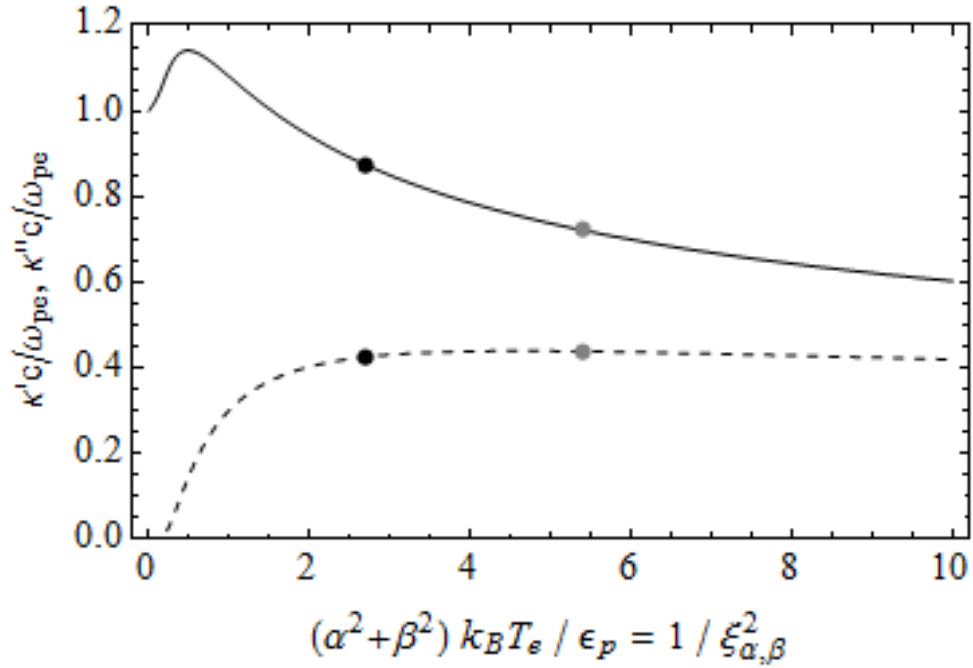

Fig. 8: Complex stochastic damping coefficient $\kappa = \kappa' + i\kappa''$. Solid line: $\kappa'$, dashed line: $\kappa''$. $\omega_{pe}/c$ is the classical collisionless damping coefficient related to an evanescent wave. Dots represent standard conditions (Appendix B): $\alpha^2 + \beta^2 = 1$ (Black), $\alpha^2 + \beta^2 = 2$ (Gray).

This is a universal relation, independent of the nature of the conductivity. The imaginary part, which leads to oscillations in space, is directly proportional to the real part of the conductivity, i.e. to the part responsible for dissipation. Therefore, oscillatory behavior and dissipation are connected generally. After inserting the stochastic conductivity from equation (30), the real and imaginary parts of the damping constant read:

$$\kappa' = \frac{\omega_{pe}}{c} \sqrt{\xi_{\alpha,\beta}} \sqrt{F(\xi_{\alpha,\beta}) + \sqrt{F^2(\xi_{\alpha,\beta}) + \frac{\pi}{4} e^{-2\xi_{\alpha,\beta}^2}}}$$

$$\kappa'' = \frac{\omega_{pe}}{c} \frac{\sqrt{\pi}}{2} \frac{\sqrt{\xi_{\alpha,\beta}} \, e^{-\xi_{\alpha,\beta}^2}}{\sqrt{F(\xi_{\alpha,\beta}) + \sqrt{F^2(\xi_{\alpha,\beta}) + \frac{\pi}{4} e^{-2\xi_{\alpha,\beta}^2}}}} \quad (38)$$

Both parts depend only linearly on the electron plasma frequency and converge to the classical values for $k_B T_e / \varepsilon_p \propto k_B T_e / \Lambda^2 \to 0$. In this respect the behavior of the real part is similar to the classical collisionless case $\left( \chi' = 1/s = \omega_{pe}/c \right)$, although in that case the imaginary part is zero. The subsequent factors are only functions of the electron temperature (Fig. 8). The imaginary part rises quickly and then becomes approximately constant within the relevant range while the real part decays slowly beyond its maximum like $(k_B T_e)^{-1/4}$. Clearly, effective stochastic heating requires temperatures $k_B T_e > \varepsilon_p$. On the other hand, temperatures will hardly exceed $5\varepsilon_p$ and within this range the effective skin depth does not increase more than 20 % above the classical value.

The first zero of the oscillatory part appears in the range between 3 to 4 skin depths. Therefore, the oscillatory behavior has little effect on most part of the electric field profile and becomes significant only when the field has decayed already substantially. The corresponding phase velocity in z-direction is $\omega_0/\kappa''$, which is an order of magnitude faster than the thermal velocity but still significantly lower than the speed of light. Overall, this justifies the simplifying assumption of an exponential decay made in the Lieberman model.

An example of the electric field decay is shown in Fig. 9. Parameters are from the standard conditions (Appendix B). Although the axial field components and the diagonal components have different damping constants, the actual difference in the field decay is small. The total field decays initially exponentially. Only at rather low field amplitudes the oscillatory part becomes significant. The logarithmic scale strongly enhances the differences but on a linear scale actually all curves are very close.

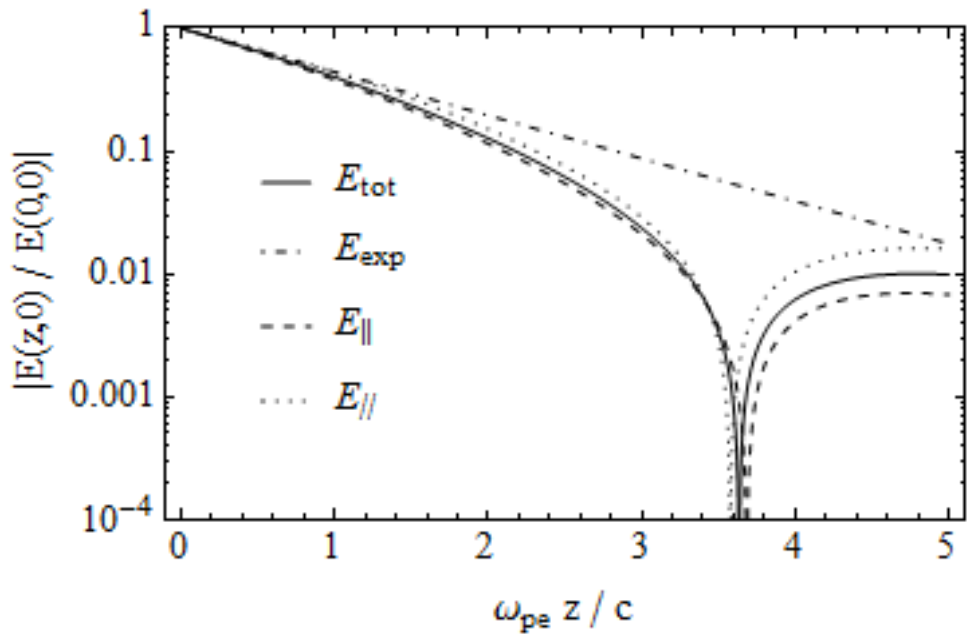

Fig. 9: Decay of the absolute value of the electric field (at $t=0$) along the vertical coordinate $z$ (distance to the antenna) for standard conditions (Appendix B). All fields are normalized to the value at $z=0$. $c/\omega_{pe}$ is the classical skin depth. The decay is slightly different for field components parallel to the axes $\left(E_{\parallel}\right)$ and diagonal components $\left(E_{//}\right)$. The total normalized field is $E_{tot}=2/3\left(E_{\parallel}+E_{//}/2\right)$. For comparison, the exponential decay of total field by only the real part of the damping coefficient is shown too $\left(E_{exp}\right)$.

### 3.3 Accounting for Elastic Collisions

Momentum transfer by collisions with neutral atoms can be accounted for by extending the Boltzmann equation (24) by a relaxation (Krook) operator on the right hand side [13, 14]: $\partial f / \partial t|_{col} = -v_m(f - f_0) = -v_m f_1$, where $v_m = $ const. is the elastic collision frequency for electron momentum exchange with background atoms. The linearized equation then reads:

$$\frac{\partial f_1}{\partial t} + v_m f_1 + \vec{v} \cdot \nabla_r f_1 \approx \frac{e}{m_e} \vec{E} \cdot \nabla_v f_0(v) \quad (39)$$

After Fourier transformation the only change compared to the above collisionless case is $\omega_0 \to \omega_0(1 - i v_m / \omega_0) = \omega_0(1 - i \zeta_m)$. Therefore, now the argument in the plasma dispersion function determining the behavior of the conductivity becomes complex. The total conductivity is:

$$\sigma_{tot}(\xi_{\alpha,\beta}, \zeta_m) = i \sigma_0 \xi_{\alpha,\beta} Z^*\left((1 - i \zeta_m)\xi_{\alpha,\beta}\right) \quad (40)$$

Although the general behavior is complex, some special cases have simple limits. In this context it is useful to recall that $\zeta_m \xi_{\alpha,\beta} = \Lambda/(2\pi \lambda_m)$, where $\lambda_m = v_{th}/v_m$ ($v_{th}$: thermal speed) is the electron mean free path, and $\xi_{\alpha,\beta}\sqrt{\alpha^2 + \beta^2} = \Lambda/(T_{RF} v_{th})$. For large arguments, the plasma dispersion function converges asymptotically to $Z(z) \to -1/z$ [12]. In this limit the conductivity becomes $\sigma_{tot} \to \sigma_0/(\zeta_m + i)$, which recovers exactly the Ohmic conductivity (equation (20)). Naturally, the Ohmic relation applies either in the case of large collisionality $\zeta_m \to \infty$ or locality $\xi_{\alpha,\beta} \to \infty$, i.e. infinite scale length of the array or vanishing electron temperature. In the stochastic case $\xi_{\alpha,\beta} \to 0$, the approximate conductivity (expansion up to second order in $\xi_{\alpha,\beta}$) reads:

$$\sigma_{tot}(\xi_{\alpha,\beta} \ll 1, \zeta_m) \approx \sigma_0 \xi_{\alpha,\beta} \left(\sqrt{\pi} - 2\zeta_m \xi_{\alpha,\beta} - i 2\xi_{\alpha,\beta}\right) \quad (41)$$

Alternatively, one can expand for low collisionality, i.e. small values of $\zeta_m \xi_{\alpha,\beta}$ up to first order:

$$\sigma_{tot}(\xi_{\alpha,\beta}, \zeta_m \ll 1) \approx \sigma_{sto}(\xi_{\alpha,\beta}) - 2\left(i \sigma_{sto}(\xi_{\alpha,\beta}) + 1\right)\xi_{\alpha,\beta}^2 \zeta_m \quad (42)$$

This expansion includes the full stochastic conductivity and provides in addition to the above case also a fist order correction to the imaginary part. The condition for negligible collisionality following from the real part is $\zeta_m \xi_{\alpha,\beta} \ll \sqrt{\pi}/2$ while the imaginary part requires $\zeta_m \xi_{\alpha,\beta} \ll 1/\sqrt{\pi}$. Using the second and slightly stricter condition and $\lambda_m = \sqrt{8 k_B T_e / m_e}/v_m$ for the mean fee path, $\lambda_m/\Lambda \gg 1/(\pi\sqrt{\alpha^2 + \beta^2}) \approx 0.32$ or $0.16$, respectively, results. The condition is even more relaxed than the initial estimate $\lambda_m/\Lambda \gg 1$, which explains the good experimental performance of the array discharge even at moderate pressures where $\lambda_m/\Lambda \approx 1$ but $\zeta_m \ll 1$, which is unfavorable for Ohmic heating. The

- 21 -

dependence of the complex conductivity on the temperature parameter $\xi_{\alpha,\beta}$ and the collision parameter $\zeta_m$ is shown in Fig. 10. Note that collisions always reduce the real part of the conductivity in the stochastic regime $\xi_{\alpha,\beta}<1$, although the effect is relatively weak. Contrary, in the local regime $\xi_{\alpha,\beta}>1$ initially the real part of the conductivity increases strongly. At $\zeta_m=0$ the transition from negative to positive slope is at $\xi_{\alpha,\beta}=0.92$ which confirms again the threshold definition at $\xi_{\alpha,\beta}=1$. In this sense, there is no Ohmic contribution to the stochastic heating in the regime $\xi_{\alpha,\beta}<1$ but collisions simply reduce coherence and therefore heating. Naturally, the imaginary part (absolute value) is always reduced by collisions.

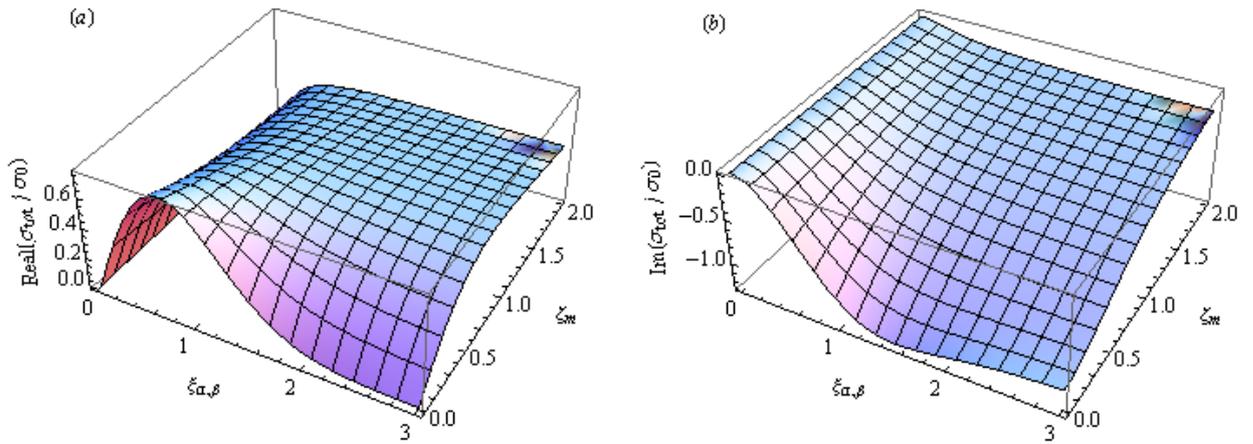

Fig. 10: Real (a) and imaginary (b) part of the total normalized conductivity $\sigma_{tot}/\sigma_0$ as a function of the temperature parameter $\xi_{\alpha,\beta}=\sqrt{\varepsilon_p/k_B T_e(\alpha^2+\beta^2)}$ and the collision parameter $\zeta_m=\nu_m/\omega_0$. The conductivity parameter is $\sigma_0=\omega_{pe}^2/(\mu_0 c^2 \omega_0)$ as defined in equation (21).

## 4. Summary and Discussion

The basic physics of collisionless heating in the INCA discharge has been investigated theoretically. It is shown that the heating is local in the vertical direction but nonlocal in the plane of the field. The anisotropic velocity distribution function, the conductivity, the mean power per area delivered to the electrons and the complex damping constant for the penetration of the RF field into the plasma are calculated. An effective stochastic collision frequency is defined by the phase shift between field and current density. Ohmic heating is taken into account and shown to have a very small contribution in the stochastic non-local regime. Three dimensionless key parameters are identified that determine the physics:

- $\chi^2$ is the ratio of the phase energy of the array $\varepsilon_p$ to thermal energy of the electrons. The parameter governs the dependence of the stochastic heating on the electron temperature. Generally, stochastic operation is within $0<\chi<1$. With proper choice of $\varepsilon_p$ the variation of all effective quantities on this parameter is rather modest.

- $\psi$ is the product of the skin depth and the wave vector of the array. The parameter controls locality. If $\psi \ll 1$ stochastic heating is dominated by non-locality along the vertical coordinate, i.e. the classical case, but is local across the plane. If $\psi \gg 1$, non-locality in the plane dominates. This is the INCA case. The skin depth is found to be the classical skin depth $c/\omega_{pe}$ multiplied by a function depending on the electron temperature only. This function is actually of the order of one and the skin depth is enlarged by not more than a few ten percent at most.
- $\zeta_m$ is the ratio of the elastic collision frequency for momentum exchange and the angular RF frequency. This parameter governs the collisionality and transition between Ohmic and stochastic heating. For $\zeta_m \chi \ll 1$ or $\lambda_m / \Lambda \gg 0.1$, collisions have a negligible influence on the complex conductivity and the stochastic processes dominate. In the opposite limit the classical Ohmic behavior is recovered.

Effectively, the vortex field in the ortho configuration can be decomposed into eight components. These components have the form of "standing waves" along the two Cartesian axes and the two diagonals. Along each of the four directions electrons can come into resonance when moving at a velocity comparable to the phase velocity of the standing wave, regardless of the sign of the velocity. In total this makes up the eight components or resonances in velocity space. The term "standing wave" is used here as an analogy motivated by the apparent similarity found for the general form of the various terms: $\sin(\vec{k} \cdot \vec{r}) \cos(\omega_0 t)$. However, it should be stressed that these are not real waves and there is no reflection anywhere. In particular, no dispersion relation connects $\omega_0$ and $k_0$ and both quantities can be chosen arbitrarily and independently. This is one of the key points of the INCA concept and allows the design to be optimized without limitations usually imposed by the dispersion relations of plasma waves. In addition, the transversal direction of the field allows energy gain without bringing the electron out of resonance, i.e. the velocity along the "standing wave" is not affected by the acceleration.

Further, stochastic heating in the INCA discharge is not related to the formation of space charge and corresponding Landau damping [13, 14]. The RF frequency is well located in the bandgap below the electron plasma frequency and above the ion plasma frequency. In addition, the electric fields of the standing waves are oriented perpendicularly to the wave vector since they originate from the induced vortex fields. Therefore, naturally no space charge is formed and electrostatic waves are not excited.

The results obtained in this work might be applied not only to the INCA discharge but also to other cases in the literature. Recently another large area discharge based on inductive coupling has been introduced by Hollenstein, Guittienne and Howling [16, 17]. This source consists of a larger number of linear rods either in a plane or in a circular arrangement. A special feature of the source is the combination of the linear inductors with capacitors at the end points so that effectively a waveguide structure is formed. The dimensions are chosen so that the circuit becomes resonant at the RF frequency and standing waves are established which allows high currents in the linear rods and efficient inductive coupling to the plasma. This setup is actually a linear version of the array discussed here. The basic spatial mode of the induced electric field is a simple sine function in the plane, i.e. $E_0 \sin(k_0 x) \vec{e}_y$, where $k_0$ is of similar order as in the present work. However, superimposed on this field structure is a macroscopic amplitude modulation due to the resonant standing wave concept, i.e. a function $\sin(Kx)$ with $K << k_0$. The spatial average over the square of this macroscopic function provides a factor ½ in the mean power density. Then the same results as derived above can be applied by considering this overall factor ½, using $E_0/2 \to E_0$. Naturally, diagonal contributions are missing and the axial terms have only a multiplicity of 2 instead of 4 since here only one axis contributes. The overall factor in the heating power (equation (33)) is then $1/2 \cdot 1/2 \cdot 2^2 = 1$, i.e. it remains unchanged, but the second exponential term in $G_{ortho}^{(V)}$ (equation (34)) has to be dropped. Overall, the behavior in terms of the physics should be quite similar to the INCA case and, indeed, data in the literature suggest that the performance is comparable [16, 17], although the technical realization is quite different.

A number of theoretical problems are still outstanding, e.g. a kinetic model predicting the actual form of the distribution function, in particular describing quantitatively the experimentally observed Maxwellisation of the distribution in case of stochastic heating. Further, the experiment shows strong indications of the formation of super-fast electrons which probably pass the array several times. These electrons lead to a significant increase in the boundary sheath potential and enhanced ionization. Clearly quantitative prediction of their fluxes needs to be addressed by a kinetic model. Further, finite size effects of the array on the electric field and the heating need to be considered in a complete model. Formally, including classical and the new stochastic heating effect in parallel in the collisionless Boltzmann equation would be desirable. Last not least effects of the induced magnetic field on the electron trajectory might be important. Although one would expect only small angle effects, still they could lead to a noticeable dephasing over larger distances. Such effects should become important at large amplitudes of the induced fields, i.e. as higher order contributions.

In conclusion, a theory for the stochastic heating in the INCA discharge has been introduced. The present theory explains the observed high efficiency of the collisionless heating. Other aspects, as outlined above, still remain the task of future work. At present development of new discharge concepts at very low pressures seems to be limited to thrusters. Outside this special field, research and application show a strong trend of shifting emphasis from low to atmospheric pressure plasmas and sources, as is for instance highlighted in the corresponding section of "The 2017 Plasma Roadmap" [18]. New concepts and ideas like the work by Hollenstein et al. and the present INCA concept hopefully will stimulate further activities in the low pressure field where certainly still new physics and new applications can be found.

**Acknowledgement**

Inspiring discussions with Tsanko V. Tsankov and Satoshi Hamaguchi are gratefully acknowledged.

## 5. References

[1] Ph. Ahr, T.V. Tsankov, J. Kuhfeld, U. Czarnetzki, *Inductively Coupled Array Discharge*, submitted to Plasma Sources Science and Technology, arXiv:1806.02043v1 (2018).

[2] U. Czarnetzki and Kh. Tarnev, *Collisionless electron heating in periodic arrays of inductively coupled plasmas*, Physics of Plasmas **21**, 123508 (2014).

[3] M. A. Lieberman and A. J. Lichtenberg, *Principles of Plasma Discharges and Materials Processing*, Wiley, New York (1994).

[4] V. Vahedi, M.A. Lieberman, G.DiPeso, T.D. Rognlien, and D. Hewett, J. Appl. Phys. **78**, 1446 (1995)

[5] M. M. Turner, *Collisionless heating in radio-frequency discharges: a review*, J. Phys. D: Appl. Phys. **42**, 194008 (2009).

[6] E. S. Weibel, *Anomalous skin effect in a plasma,* Phys. Fluids **10**, 741 (1967).

[7] S. Ichimaru *Basic principles of plasma physics: a statistical approach*, Benjamin, Reading, London (1973).

[8] Milton Abramowitz and Irene A. Stegun, *Handbook of Mathematical Functions*, Dover Publications Inc., New York (1973).

[9] V. Godyak, *Hot plasma Effects in gas discharge plasmas*, Physics of Plasmas **12**, 055501 (2005).

[10] L.C. Pitchford, L.L. Alves, K. Bartschat et al, *Comparison of sets of electron-neutral scattering cross sections and swarm parameters in noble gases: I. Argon*, Journal of Physics D: Appl. Phys. **46**, 334001 (2013).

[11] L.L. Alves, K. Bartschat, S.F. Biagi, et al., *Comparison of sets of electron-neutral scattering cross sections and swarm parameters in noble gases: II. Helium and neon*, Journal of Physics D: Appl. Phys. **46**, 334002 (2013).

[12] Burton D. Fried and Samuel D. Conte, *The Plasma Dispersion Function*, Academic Press Inc., New York 1961.

[13] T.H. Stix, *Waves in Plasmas*, American Institute of Physics, New York (1992).

[14] J.A. Bittencourt, *Fundamentals of Plasma Physics*, Springer, New York (2004).

[15] Y. Raizer, *Gas Discharge Physics*, Springer, Berlin, 1991.

[16] Ch. Hollenstein, Ph. Guittienne, and A.A. Howling, *Resonant RF network antennas for large-area and large-volume inductively coupled plasma sources*, Plasma Sources Science and Technology **22**, 055021 (2013)

[17] Ph. Guittenne, R. Jacquier, A.A. Howling, I. Furno, *Electromagnetic, complex image model of a large area RF resonant antenna as inductive plasma source*, Plasma Sources Science and Technology, **26**, 035010 (2017).

[18] I. Adamovich, S.D. Baalrud, A. Bogaerts, et al., *The 2017 Plasma Roadmap: Low temperature plasma science and technology*, Journal of Physics D: Appl. Phys. **50**, 323001 (2017).

**Appendix A: Electric Fields of the Ortho and the Para Array**

The fundamental electrical fields of the ortho and the para array in the *x,y*-plane without any *z*-dependence were first introduced in Ref. [2] and are repeated here for convenience, although with a slightly different notation. The term "fundamental" means that in the real field in the experiment also higher spatial modes are present but with much lower amplitudes. Their contribution is neglected here. The electric field for the ortho array is (amplitude $E_0$):

$$\vec{E}_{ortho} = \frac{E_0}{2} \begin{pmatrix} (1+\cos(k_0 x))\sin(k_0 y) \\ -(1+\cos(k_0 y))\sin(k_0 x) \\ 0 \end{pmatrix} \cos(\omega_0 t) \ . \tag{43}$$

Here $k_0 = 2\pi/\Lambda$ is the wave vector of the array with $\Lambda$ representing the length of the primitive cell after which the structure reproduces itself. For the ortho array this is the size of the individual coil cell. $\omega_0 = 2\pi/T_{RF}$ is the angular frequency of the applied harmonic RF current with period $T_{RF}$. This compact presentation can be decomposed into 8 vector components, with 6 non-vanishing terms per each of x,y-coordinate directions:

$$\vec{E}_{ortho} = \frac{E_0}{8i} \sum_{\alpha,\beta=0,\pm 1} \vec{A}_{\alpha,\beta}\, e^{i(\vec{k}_{\alpha,\beta}\cdot\vec{r}+\omega_0 t)} + c.c. \tag{44}$$

The vector amplitudes of the electric field and the corresponding wave vectors read:

$$\vec{A}_{\alpha,\beta} = \frac{1}{\alpha^2+\beta^2}\begin{pmatrix}\beta\\-\alpha\\0\end{pmatrix} \quad \text{and} \quad \vec{k}_{\alpha,\beta} = k_0\begin{pmatrix}\alpha\\\beta\\0\end{pmatrix} \tag{45}$$

The field components with one index being zero represent fields along the coordinate axes while components with both indices non-zero represent fields along the diagonals.

The fundamental electrical field of the para array in the *x,y*-plane without the *z*-dependence is:

$$\vec{E}_{para} = E_0 \begin{pmatrix} \sin(k_0 x)\cos(k_0 y) \\ -\cos(k_0 x)\sin(k_0 y) \\ 0 \end{pmatrix} \cos(\omega_0 t) \ . \tag{46}$$

In the corresponding vector components, terms with $\alpha,\beta=0$ are missing, i.e. field terms along the coordinate axes, and the remaining amplitudes are a factor of 2 larger than for the ortho array:

$$\vec{E}_{para} = \frac{E_0}{4i} \sum_{\alpha,\beta=\pm 1} \vec{A}_{\alpha,\beta}\, e^{i(\vec{k}_{\alpha,\beta}\cdot\vec{r}+\omega_0 t)} + c.c. \tag{47}$$

Note that for both array structures some general properties apply that are important for the calculation in this work: The corresponding wave vectors are perpendicular to the field vectors as required for a divergence free vortex field: $\vec{A}_{\alpha,\beta}\cdot\vec{k}_{\alpha,\beta}=0$. Further, $\left|\vec{A}_{\alpha,\beta}\right|^2 = 1/\left(\alpha^2 + \beta^2\right)$ and $\left|\vec{k}_{\alpha,\beta}\right| = k_0\sqrt{\alpha^2 + \beta^2}$, where for the ortho array $\alpha^2 + \beta^2 = 1$ or $2$, respectively, and $\alpha^2 + \beta^2 = 2$ for the para array.

**Appendix B: Standard Experimental Conditions**

The experimental standard conditions used for calculation of explicit numerical values are taken from Ref. [1]. All numerical values are rounded to 2 digit resolution which is sufficient within the context of this work and also characteristic for most experimental values.

Working gas: Argon

Pressure: $p = 1\,\text{Pa}$

Angular RF frequency: $\omega_0 = 0.85 \cdot 10^8\,\text{s}^{-1}$

RF power (power actually delivered to the plasma, and coil current): $P_{RF} = 430\,\text{W}$, $I = 12\,\text{A}$

Cell size, number of cells (coils): $\Lambda = 0.050\,\text{m}$, $N \times N = 6 \times 6 = 36$

Characteristic array wave vector: $k_0 = 2\pi/\Lambda = 130\,m^{-1}$

Array wave vectors $(\alpha, \beta = 0, \pm 1)$: $\vec{k}_{\alpha,\beta}/k_0 = (\alpha, \beta, 0)$

Antenna area: $A = (N\Lambda)^2 = 0.090\,\text{m}^2$

Measured electron temperature ($k_B$: Boltzmann constant, $e$: elementary charge):
$k_B T_e / e = 3.5\,\text{eV}$

Skin depth: $s = 0.032\,m$

Mean electron density within the skin depth: $n_0 = 3.0 \cdot 10^{16}\,m^{-3}$

Plasma frequency: $\omega_{pe} = \sqrt{\dfrac{e^2 n_0}{\varepsilon_0 m_e}} = 9.8 \cdot 10^9\,s^{-1}$

($\varepsilon_0$: vacuum permittivity, $m_e$: electron mass)

Elastic collision frequency $(p = 1\,Pa, T_{gas} = 400\,K)$ [15]: $\nu_m = 3.0 \cdot 10^7\,s^{-1}\,Pa^{-1}$

Array phase velocity: $v_p = \omega_0 / k_0 = 0.68 \cdot 10^6\,m/s$

Array phase energy: $\varepsilon_p / e = m_e v_p^2 / (2e) = 1.3\,eV$

Dimensionless parameters:

$\psi_0 = s\,k_0 = 4.2$, $\chi_0 = \sqrt{\varepsilon_p/(k_B T_e)} = 0.61$,

$\zeta_m = \nu_m/\omega_0 = 0.35$, $\alpha, \beta = 0, \pm 1$, $\sqrt{\alpha^2 + \beta^2} = 1$ or $\sqrt{2}$,

$\xi_{\alpha,\beta} = \chi_0/\sqrt{\alpha^2 + \beta^2} = 0.61$ or $0.43$

**Appendix C: Analytical Approximations for the Integrals $G_1$ and $G_2$**

The two integral functions appearing in this work, $G_1$ and $G_2$, are both containing in the integrand the function $H(x^2)$, defined by equation (11). This function has properties similar to a $\delta(x)$-function, i.e. it diverges at zero, has a finite area, and is very narrow, although not infinitely narrow (Fig. 10). As will be shown below, in the ultimate limit the function $\lim_{\psi \to \infty} \psi H((\psi x)^2)$ would converge exactly to the delta function. Here, the actual value of $\psi$, appearing in the integrals defining $G_{1,2}$, is finite and the delta function properties are realized only approximately. A discussion of the related error is also given below. The final conclusion of this analysis is that one can replace with good accuracy the $H$ function by a $\delta$-function when integrating, i.e. for $\psi^2 \gg 1/2$:

$$\psi H\left((\psi x)^2\right) \approx \frac{\pi^{3/2}}{2}\delta(x) \tag{48}$$

The required property for a $\delta$-function is (arbitrary function $g(x)$ and constant $C$):

$$g(x) \overset{!}{=} \lim_{\psi \to \infty} \psi C \int_{-\infty}^{+\infty} H\left((\psi(u-x))^2\right) g(u)\,du$$
$$= \lim_{\psi \to \infty} C \int_{-\infty}^{+\infty} H(w^2) g\left(\frac{w}{\psi}+x\right) dw \tag{49}$$
$$= g(x) C \int_{-\infty}^{+\infty} H(w^2)\,dw$$

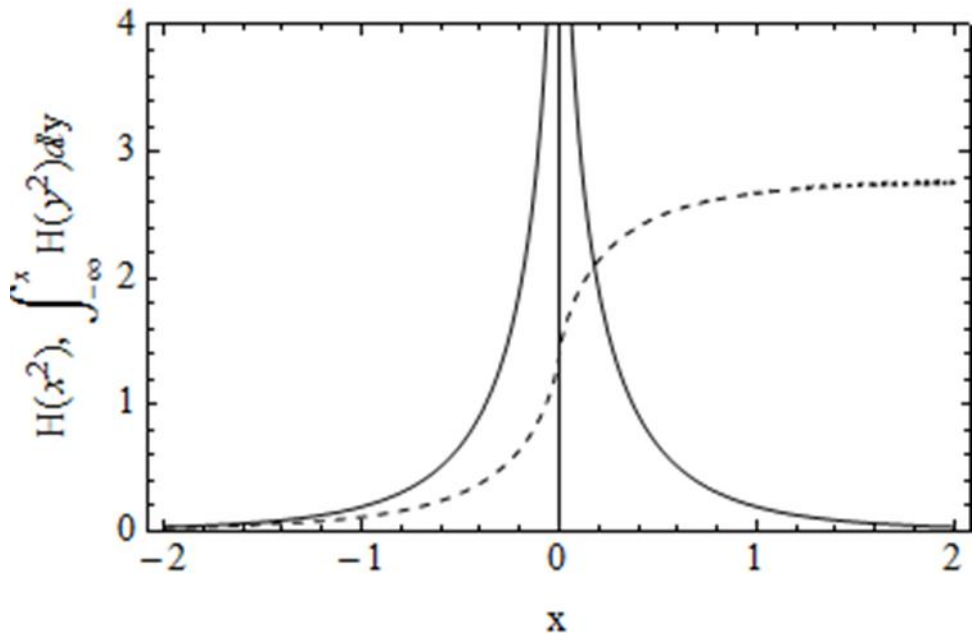

Fig. 10: The function $H(x^2)$ (solid line) and its integral starting integration at minus infinity (dashed line).

The integral determining the value of $C$ is evaluated by taking advantage of the integral definition of the function $H$ (equation (11)) and exchanging the order of integration:

$$\int_{-\infty}^{\infty} H(w^2) dw = \int_0^{\infty} y e^{-y} \int_{-\infty}^{\infty} \frac{1}{(y+w^2)^2} dw\, dy$$

$$= \int_0^{\infty} \frac{e^{-y}}{\sqrt{y}} dy \int_{-\infty}^{\infty} \frac{1}{(1+w^2)^2} dw \qquad (50)$$

$$= 2 \int_0^{\infty} e^{-y^2} dy \frac{\pi}{2}$$

$$= \frac{\pi^{3/2}}{2}$$

Therefore, with $C = 2/\pi^{3/2}$ the required $\delta$-function properties are indeed met for infinite values $\psi \to \infty$. For finite values of $\psi$, the function $g(w/\psi + x)$ can be expanded within the integral (equation (49)) to second order around $x$. Naturally, all odd terms vanish in the integral. A similar integration procedure as in equation (50) yields for the right hand side:

$$g(x)\left(1 + \frac{1}{4\psi^2} \frac{g''(x)}{g(x)}\right) = g(x)\left(1 + \frac{2x^2 - 1}{2\psi^2}\right) \qquad (51)$$

The explicit expression on the right follows for the particular case of $g(x) \sim \exp(-x^2)$ found in the $G_1$ integral. A negligible relative error requires $\psi^2 \gg |2x^2 - 1|/2$. Since here $x = \chi$ and for all reasonable parameters $\chi \leq 1$ (or $k_B T_e \geq \varepsilon_p$), the final requirement is $\psi^2 \gg 1/2$ (or $s \gg 0.11 \Lambda$). This condition is indeed well met (standard conditions: $\psi_0^2 \approx 18$). Finally, $x = \chi/\sqrt{2}$ in case of $G_2$ leads to even slightly more relaxed numerical values so that the above conclusion applies generally. With the approximate $\delta$-function property of $H$ confirmed, the integrals for $G_{1,2}$ can be evaluated immediately:

$$G_1(\chi) = \frac{\psi}{\sqrt{\pi}} \int_{-\infty}^{\infty} H\left(\psi^2 (u-\chi)^2\right) e^{-u^2} du \approx \frac{\pi}{2} e^{-\chi^2} \qquad (52)$$

Similarly one finds:

$$G_2(\chi) = \frac{\psi}{\sqrt{2\pi}} \int_{-\infty}^{\infty} H\left(\psi^2 (u-\chi)^2\right) e^{-\frac{1}{2} u^2} du \approx \frac{\pi}{2^{3/2}} e^{-\chi^2/2} \qquad (53)$$

Fig. 11 shows that there is in fact excellent agreement between these approximate analytical expressions and exact numerical integration. In addition, the vanishing error for $G_1$ at $1/\chi^2 = k_B T_e / \varepsilon_B = 2$, the sign of the error, the maximum error value of 4 % at large electron temperatures $(\chi \to 0)$, and the smaller error for $G_2$ are confirmed. This clearly highlights the rapid convergence of the expansion in $\psi$ and the quality of the $\delta$-function approximation.

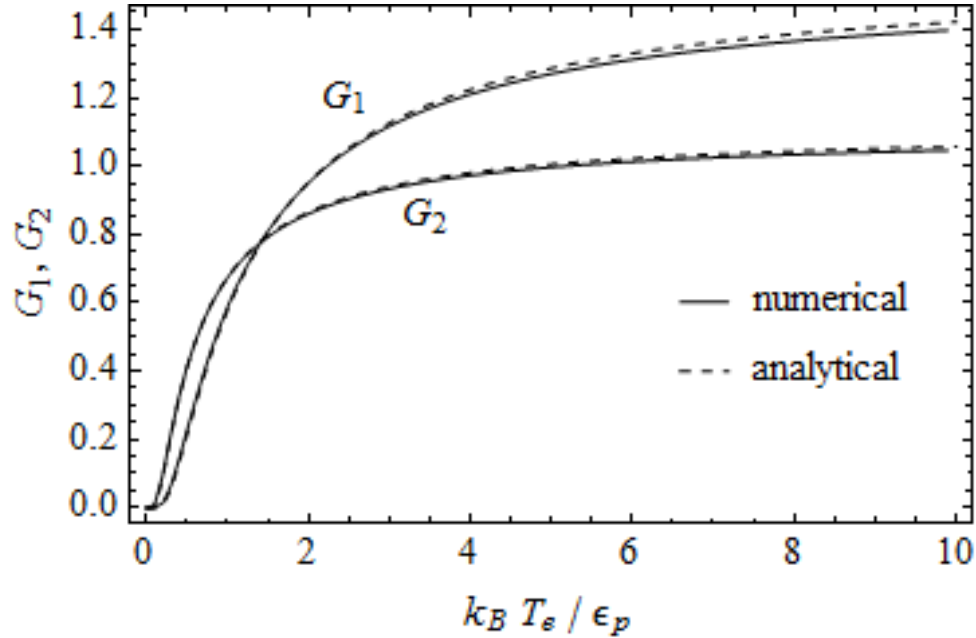

Fig. 11: Comparison between the numerical integration (solid lines) and the analytical approximation (dashed lines) for the functions $G_1$ and $G_2$. The numerical integration is carried out for $\psi_0 = 4.2$ (standard conditions).

In the opposite limit $\psi \ll 1$ but with $\chi\psi$ remaining finite, the integrals converge again to analytical expressions:

$$\lim_{\psi \to 0} \chi G_1(\chi) = \lim_{\psi \to 0} \frac{\chi\psi}{\sqrt{\pi}} \int_{-\infty}^{\infty} H\left(\psi^2(u-\chi)^2\right) e^{-u^2} du$$
$$= \lim_{\psi \to 0} \frac{\chi\psi}{\sqrt{\pi}} H\left((\psi\chi)^2\right) \int_{-\infty}^{\infty} e^{-u^2} du \qquad (54)$$
$$= \chi\psi H\left((\psi\chi)^2\right)$$

Exactly the same result is found for $\chi G_2$ so that finally:

$$\lim_{\psi \to 0} G_{ortho}^{(L)} = \frac{3}{2} \chi\psi H\left((\psi\chi)^2\right) \qquad (55)$$

This limit is found in the case of classical stochastic heating which is recovered when $\Lambda \to \infty$. The product $\chi\psi$ remains finite since there $\Lambda$ cancels out and the skin depth $s$ becomes the relevant scaling length.